\documentclass[twocolumn,amsmath,pra,superscriptaddress,nobibnotes,showpacs]{revtex4}
\usepackage{graphicx}

\begin{document}
\title{Ghost Interference with Optical Parametric Amplifier}

\author{Sulakshana Thanvanthri}
\email{su1@umbc.edu} \affiliation{Department of Physics,
University of Maryland, Baltimore County, Baltimore, Maryland,
21250}

\author{Morton H. Rubin}
\affiliation{Department of Physics, University of Maryland,
Baltimore County, Baltimore, Maryland, 21250}
\date{\today}

\begin{abstract} The 'Ghost' interference experiment is analyzed
when the source of entangled photons is a multimode Optical
parametric Amplifier(OPA) whose weak limit is the two-photon
Spontaneous Parametric Downconversion(SPDC) beam. The visibility
of the double-slit pattern is calculated, taking the finite
coincidence time window of the photon counting detectors into
account. It is found that the coincidence window and the bandwidth
of light reaching the detectors play a crucial role in the loss of
visibility on coincidence detection, not only in the 'Ghost'
interference experiment but in all experiments involving
coincidence detection. The differences between the loss of
visibility with  two-mode and multimode OPA sources is also
discussed.
\end{abstract}
\maketitle
\section{Introduction}

 The 'Ghost' interference experiment is typical of
two-photon interference experiments that bring out quantum
entanglement features of light. These features are not just
confined to the appearance of the double-slit pattern on
coincidence detection. The independence of the result on where the
optical elements are situated in the experimental set-up (a
double-slit in the case of 'Ghost' interference) is a key feature
of entanglement. The 'Ghost' interference experiment has been
performed in the low-gain limit of parametric
down-conversion~\cite{shih}. It is of interest to examine this
experiment in the high gain regime of parametric amplification.
Analysis of similar experiments using two-mode OPA
states~\cite{boyd} suggest the loss of visibility at large gains.
Recently different detection schemes have been proposed to
circumvent this problem ~\cite{gatti1, gatti2}. Here we present a
detailed calculation of the 'Ghost' interference experiment using
a multimode OPA as the source of entangled photons. We analyze the
effect of the coincidence time window of the photon counting
detectors on the experimentally observable interference pattern
and visibility. We attempt to explain how the properties of the
source of entangled photons and experimental limitations affect
the observable interferometric effects. We find that the loss of
visibility depends on the coincidence time window and the
bandwidth of the source. The coincidence time window causes a loss
in visibility at much lower gains than expected with ideal
detectors. A multimode source causes loss in visibility even if
ideal detectors are used. Combined with a finite coincidence time
window, a multimode source can reduce visibility to less than 0.5
even at very low parametric gain. These effects are not specific
to the 'Ghost' interference experiment but occur in all
experiments with similar sources and detection schemes.
\vspace{0.5in}

\section{Multimode interaction in OPA}
We consider a non-degenerate OPA~\cite{mollow} comprising a
non-centrosymmetric crystal with a second order non-linear
susceptibility $\chi^{(2)}$. The crystal is pumped by a CW laser,
given by
\begin{eqnarray}\label{pump}
    E_{p}(\vec{r},t) & = & E_{p}^{(+)}+ E_{p}^{(-)}\\
    & = & E_{p} e^{i (\vec{k}_{p} \cdot \vec{r} - \omega_{p}
    t)}+ H.c
\end{eqnarray}
where H.c stands for Hermitian conjugate. We assume that the pump
is not depleted, has a constant amplitude $E_{p}$ and can be
treated classically. If a weak multimode signal is also input into
the OPA, the output after parametric interaction within the
crystal consists of amplified modes of the signal and idler where
the idler modes obey the phase matching conditions,
\begin{eqnarray}\label{phase-match}
\omega_{p} & = & \omega_{s}+\omega_{i} \\
\vec{k}_{p} & = & \vec{k}_{s} + \vec{k}_{i}
\end{eqnarray}
where $\vec{k}_{p}$, $\vec{k}_{s}$ and $\vec{k}_{i}$ are the wave
vectors of the pump, signal and idler photons.
 The quantized field of the signal and idler
inside the crystal are given by ~\cite{glauber},
\begin{eqnarray}\label{signal-idler}
    E_{s}(\vec{r},t) & = &
    \sum_{k_{s}}\sqrt{\frac{\hbar \omega_{s}}{2\epsilon_{0}V}} e^{i (\vec{k}_{s} \cdot \vec{r} - \omega_{s}
    t)}a_{\vec{k}_{s}} + H.c \\
E_{i}(\vec{r},t) & = &
    \sum_{k_{s}}\sqrt{\frac{\hbar \omega_{i}}{2\epsilon_{0}V}}e^{i (\vec{k}_{i} \cdot \vec{r} - \omega_{i}
    t)}a_{\vec{k}_{i}} + H.c
\end{eqnarray}
Here $a_{\vec{k}}$ is the annihilation operator for a photon in
mode $\vec{k}$. Note that the dispersion relation $\omega(k)$ is
different from that in vacuum, $\omega(k)\neq ck$.

The non-linear interaction between the crystal, the pump, signal
and idler modes are characterized by the interaction
Hamiltonian~\cite{klyshko}~\cite{louisell} ,
\begin{eqnarray}\label{Hint}
    H_{int} & = & 2\epsilon_{0}\chi^{(2)} \int d^{3}r E_{p}^{(+)}
    E_{s}^{(-)}E_{i}^{(-)}+ H.c\nonumber\\
    & = & 2\epsilon_{0}\chi^{(2)}  \sum_{k_{s}, k_{i}}\sqrt{\frac{\hbar \omega_{s}}{2\epsilon_{0}V}}
    \sqrt{\frac{\hbar \omega_{i}}{2\epsilon_{0}V}}E_{p} \times \nonumber\\
    & & \int d^{3}r  [e^{i (\vec{k}_{p}-\vec{k}_{s}-\vec{k}_{i}) \cdot \vec{r}}
     e^{i(\omega_{s}+\omega_{i}-\omega_{p})t}a^{\dagger}_{\vec{k}_{s}}a^{\dagger}_{\vec{k}_{i}}\nonumber\\
    & & + e^{-i(\vec{k}_{p}-\vec{k}_{s}-\vec{k}_{i}) \cdot\vec{r}}
     e^{-i(\omega_{s}+\omega_{i}-\omega_{p})t}a_{\vec{k}_{s}}a_{\vec{k}_{i}}]
\end{eqnarray}
where the integration is over the volume of the crystal
illuminated by the pump. For a long crystal and a wide pump beam,
\begin{equation}\label{large-crystal}
     \int d^{3}r e^{i (\vec{k}_{p}-\vec{k}_{s}-\vec{k}_{i}) \cdot
     \vec{r}} =  V \delta_{\vec{k}_{p}-\vec{k}_{s}, \vec{k}_{i}}
\end{equation}
 The interaction Hamiltonian is simplified to
\begin{eqnarray}\label{Hint-simplified}
H_{int} & = & 2\epsilon_{0}\chi^{(2)}V\sum_{k_{s},
k_{i}}\sqrt{\frac{\hbar\omega_{s}}{2\epsilon_{0}V}}
\sqrt{\frac{\hbar\omega_{i}}{2\epsilon_{0}V}}E_{p}\delta_{\vec{k}_{p}-\vec{k}_{s}, \vec{k}_{i}}\\
 \nonumber &
&\left[e^{i(\omega_{s}+\omega_{i}-\omega_{p})t}a^{\dagger}_{\vec{k}_{s}}a^{\dagger}_{\vec{k}_{i}}
     + e^{-i(\omega_{s}+\omega_{i}-\omega_{p})t}a_{\vec{k}_{s}}a_{\vec{k}_{i}}\right]
\end{eqnarray}
 The delta function in Eq.~(\ref{Hint-simplified}) indicates the entanglement of the OPA states
  in wave vector space. Combining the photon commutation relations and the time evolution equations for
 signal and idler modes, the time-evolved signal and idler are given by~\cite{scully,walls},
\begin{eqnarray}
\label{time-evol-signal-2} a_{\vec{k}_{s}}(t) & = &
a_{\vec{k}_{s}}(0) cosh(|\xi_{k_{s}}|) - i
a^{\dagger}_{\vec{k}_{i}}(0) sinh(|\xi_{k_{s}}|) \\
\label{time-evol-idler-2} a^{\dagger}_{\vec{k}_{i}}(t) & = &
a^{\dagger}_{\vec{k}_{i}}(0)cosh(|\xi_{k_{s}}|)+ i
a_{\vec{k}_{s}}(0)sinh(|\xi_{k_{s}}|)
\end{eqnarray}
where $t$ is the average time taken by the photons to cross the
crystal. $ a_{\vec{k}_{s}}(0)$ and $a^{\dagger}_{\vec{k}_{i}}(0)$
are the annihilation and creation operators for the input signal
and idler mode. The factors $cosh(|\xi_{k_{s}}|)$ and
$sinh(|\xi_{k_{s}}|)$ are the amplification factors that depend on
the strength of the non-linearity $\chi^{(2)}$,the pump $E_{p}$
and the frequency of the signal and idler modes. Perfect
phase-matching ensures that each signal mode interacts with only
one idler mode, selected by the phase-matching conditions.
Eq's.~(\ref{time-evol-signal-2}) and (\ref{time-evol-idler-2})
relate the photon creation annihilation operators at the output
face of the crystal to those at the input. They are derived from
the unitary transformation,
\begin{equation}\label{unitary-op}
    U_{k_{i}} =
    e^{\left[\frac{\xi_{k_{s}}^{\ast}}{2}a_{\vec{k}_{s}}(0)a_{\vec{k}_{i}}(0)
    -\frac{\xi_{k_{s}}}{2}a^{\dagger}_{\vec{k}_{s}}(0)a^{\dagger}_{\vec{k}_{i}}(0)\right]}
\end{equation}

Here $\xi_{k_{s}} = (\chi^{(2)}E_{p}\sqrt{\omega_{s}\omega_{i}}t)
e^{i\pi/2}$, may be recognized as the squeeze parameter of the
OPA~\cite{scully}. The state at the output of the crystal is given
by
\begin{equation}\label{output-state}
    |\psi\rangle = e^{-i \left(\frac{H_{int}}{\hbar}t\right)}|0\rangle
\end{equation}
which is a multiphoton state. In the limit $|\xi_{k_{s}}| \ll 1$,
the expansion of Eq.~(\ref{output-state}) can be limited to first
order in the interaction Hamiltonian, giving a vacuum state and an
entangled two-photon state. This is the two-photon limit of SPDC.

\section{Experimental Set-up and Calculation}
\begin{figure}[h]
\label{set-up}
\includegraphics [width=3.2in]{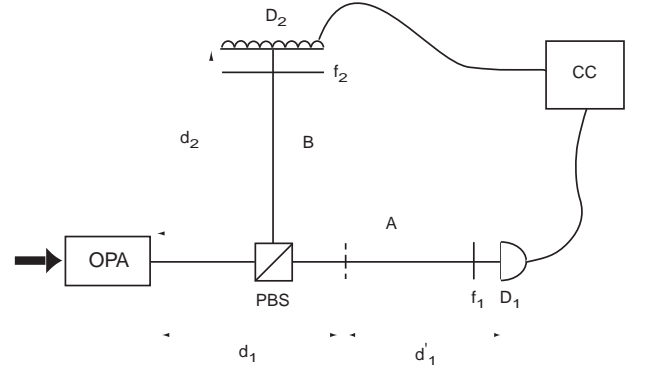}
\caption{\protect Schematic of the experimental setup for
observing the two-photon 'Ghost' interference. For simplicity, the
prism to remove the pump is not shown. The double-slit has width
$b=0.165$ mm and slit distance $a = 0.4$ mm. The relevant
distances in this experiment are $d_{1}=0.3$ m, $d_{2}=1.5$ m, and
$d^{'}_{1}=1$ m.}
\end{figure}

 We now look at the experimental set-up of the ghost
interference experiment~\cite{shih}. A non-linear crystal is
pumped by a CW laser($\lambda_{p}$ = 351nm), to generate pairs of
collinear, orthogonally polarized signal(e-ray) and idler(o-ray)
photons (Type-II SPDC). The signal and idler beams are separated
by a polarizing beam splitter. The signal beam passes through a
double slit aperture to a photon counting detector $D_{1}$.
$D_{1}$ is a fixed point detector. The idler is scanned by an
optical fiber and the output from the fiber is coupled to a
detector ($D_{2}$). $f_{1}$ and $f_{2}$  are filters (spatial or
spectral) that limit the number of wavevectors (and bandwidth) of
light reaching the detectors. The detectors in the signal and
idler arms are connected to a coincidence circuit. 
\section{Coincidence Detection}
 We are interested in coincident detection from the
signal and idler when the input state is a vacuum. The probability
of detecting a photon in $D_{1}$ in position $\vec{r}_{1}$ at time
$t_{1}$ and another in $D_{2}$ at $\vec{r}_{2}$ and $t_{2}$ is
proportional to the second order correlation
function~\cite{glauber}, given by,

\begin{equation}\label{A12}
      G^{(2)}(\vec{r}_{1},t_{1},\vec{r}_{2},t_{2})
        =  \langle \Psi | \Psi \rangle   \\
\end{equation}
\begin{equation}
      | \Psi \rangle =   E_{1}^{(+)}(\vec{r}_{1},t_{1})
   E_{2}^{(+)}(\vec{r}_{2},t_{2})  |0\rangle
\end{equation}
where $E_{1}(\vec{r}_{1},t_{1})$ and $E_{2}(\vec{r}_{2},t_{2})$
are the electric fields at the detectors $D_{1}$ and $D_{2}$. The
fields at the detectors can be expressed in terms of the fields at
the output face of the crystal through Green's functions which
describe the propagation of the beams through the optical system.
The positive frequency part of the fields at the detectors are
then given by
\begin{eqnarray}
\label{detector-fields-1}
    E_{1}^{(+)}(\vec{r}_{1}, t_{1}) & = & \sum_{k_{o}}\sqrt{\frac{\hbar \omega_{o}}{2\epsilon_{0}V}}
    e^{-i \omega_{o}t_{1}}
    g_{A}(\vec{k}_{o},\vec{r}_{1})
    a_{k_{o}} \\
    \label{detector-fields-2}
E_{2}^{(+)}(\vec{r}_{2}, t_{2}) & = &
\sum_{k_{e}}\sqrt{\frac{\hbar \omega_{e}}{2\epsilon_{0}V}} e^{-i
\omega_{e}t_{2}}g_{B}(\vec{k}_{e},\vec{r}_{2}) a_{k_{e}}
\end{eqnarray}
We now look at the state of the fields at the two detectors when
the input state is a vacuum. Note that we are relating the fields
at the detectors to the vacuum at the input face of the crystal.
This involves two transformations - from the detectors to the
output face of the crystal through the Green's functions and from
the output of the crystal to its input by the unitary
transformation of the operators in
Eq's.~(\ref{time-evol-signal-2}) and (\ref{time-evol-idler-2}).

 Using the operator commutation relations we get,
\vspace{-0.095in}
\renewcommand{\arraystretch}{2.0}
\begin{equation}
\label{state-2-terms}
\begin{array}{ll}
E_{1}^{(+)}E_{2}^{(+)}|0\rangle = & \\
\sum_{k^{'}_{o},k_{e}} -i \sqrt{\frac{\hbar
\omega^{'}_{o}}{2\epsilon_{0}V}} \sqrt{\frac{\hbar
\omega_{e}}{2\epsilon_{0}V}}
e^{-i(\omega^{'}_{o}t_{1})}e^{-i\omega_{e}t_{2}}
g_{A}(\vec{k}^{'}_{o},\vec{r}_{1}) &  \\  \times
g_{B}(\vec{k}_{e},\vec{r}_{2})cosh(|\xi_{k^{'}_{o}}|)
sinh(|\xi_{k_{o}}|) \delta_{k^{'}_{o}k_{o}}|0\rangle  - &  \\
\nonumber \sum_{k^{"}_{o},k_{e}}\sqrt{\frac{\hbar
\omega^{"}_{o}}{2\epsilon_{0}V}}\sqrt{\frac{\hbar
\omega_{e}}{2\epsilon_{0}V}}
e^{-i\omega^{"}_{o}t_{1}}e^{-i\omega_{e}t_{2}}
g_{A}(\vec{k}^{"}_{o},\vec{r}_{1}) & \\
g_{B}(\vec{k}_{e},\vec{r}_{2})sinh(|\xi_{k^{"}_{o}}|)sinh(|\xi_{k_{o}}|)
a^{\dagger}_{k^{"}_{e}}(0)a^{\dagger}_{k_{o}}(0)|0\rangle &
\end{array}
\end{equation}
 From a careful examination of the above equation,
we find that the first term indicates correlation between the
modes detected in the two arms. The second term has no such
correlation since it factors into two independent terms, one for
each arm. Physically, this term corresponds to accidental
coincidences of photons that are not entangled with each other. We
expect that the double slit diffraction pattern will emerge from
the correlation in the first term while the second term causes a
loss in visibility. The second order correlation function
$G^{(2)}$ can now be written as, \vspace{-0.1in}
\begin{equation}
\begin{array}{lll}
\label{f-corr}
G^{(2)} &  =  & \nonumber\\
& & \left|\sum_{k^{'}_{o},k_{e}}\sqrt{\frac{\hbar
\omega^{'}_{o}}{2\epsilon_{0}V}}\sqrt{\frac{\hbar
\omega_{e}}{2\epsilon_{0}V}}e^{-i
\omega^{'}_{o}t_{1}}e^{-i\omega_{e}t_{2}}
g_{A}(\vec{k}^{'}_{o},\vec{r}_{1})\right. \nonumber\\& & \left.
g_{B}(\vec{k}_{e},\vec{r}_{2})cosh(|\xi_{k^{'}_{o}}|)
sinh(|\xi_{k_{o}}|) \delta_{k^{'}_{o}k_{o}}\right|^{2}
\nonumber\\& &  + \sum_{k^{'}_{o}, k^{"}_{o}} \sqrt{\frac{\hbar
\omega^{'}_{o}}{2\epsilon_{0}V}} \sqrt{\frac{\hbar
\omega^{"}_{o}}{2\epsilon_{0}V}} e^{i\omega^{'}_{o}t_{1}}
e^{-i\omega^{"}_{o}t_{1}}
g^{\ast}_{A}(\vec{k}^{'}_{o},\vec{r}_{1})\nonumber \nonumber\\& &
g_{A}(\vec{k}^{"}_{o},\vec{r}_{1})
sinh(|\xi_{k^{'}_{o}}|)sinh(|\xi_{k^{"}_{o}}|)\delta_{k^{'}_{o}k^{"}_{o}}
\nonumber\\ & & \times \sum_{\tilde{k}_{e},
k_{e}}\sqrt{\frac{\hbar\tilde{\omega}_{e}}{2\epsilon_{0}V}}
\sqrt{\frac{\hbar\omega_{e}}{2\epsilon_{0}V}}
e^{i\tilde{\omega}_{e}t_{2}}e^{-i\omega_{e}t_{2}}
g^{\ast}_{B}(\tilde{\vec{k}}_{e},\vec{r}_{2}) \nonumber\\ & &
g_{B}(\vec{k}_{e},\vec{r}_{2})
sinh(|\xi_{\tilde{k}_{o}}|)sinh(|\xi_{k_{o}}|)\delta_{\tilde{k}_{e}k_{e}}
\end{array}
\end{equation}
 Implementing all the $\delta$ functions and converting the
summations in $G^{(2)}$ to
integrals~\footnote{$\sum_{k}\rightarrow \frac{V}{(2\pi)^{3}}\int
d^{3}k$} , we have
\begin{equation}
\begin{array}{ll}
\label{f-corr-int} G^{(2)} =&
\left(\frac{V}{(2\pi)^{3}}\right)^{2}\times \\
&\left[\left|\int d^{3}k_{e}\sqrt{\frac{\hbar
(\omega_{p}-\omega_{e})}{2\epsilon_{0}V}}\sqrt{\frac{\hbar
\omega_{e}}{2\epsilon_{0}V}}e^{-i
(\omega_{p}-\omega_{e})t_{1}}e^{-i\omega_{e}t_{2}}
\right.\right.\\& \left.\left.
g_{A}(\vec{k}_{p}-\vec{k}_{e},\vec{r}_{1})
g_{B}(\vec{k}_{e},\vec{r}_{2})
cosh(|\xi_{k_{o}}|)sinh(|\xi_{k_{o}}|) \right|^{2} \right. \\& +
\left.\int d^{3}k^{'}_{o}\frac{\hbar
\omega^{'}_{o}}{2\epsilon_{0}V}
|g_{A}(\vec{k}^{'}_{o},\vec{r}_{1})|^{2}
sinh^{2}(|\xi_{k^{'}_{o}}|)\right.\\ &\left.\times \int d^{3}k_{e}
\frac{\hbar\omega_{e}}{2\epsilon_{0}V}
|g_{B}(\vec{k}_{e},\vec{r}_{2})|^{2}
sinh^{2}(|\xi_{k_{o}}|)\right]
\end{array}
\end{equation}
The expression for $G^{(2)}$ in Eq~(\ref{f-corr-int}) can be
rewritten  as follows to emphasize the nature of the correlation
in each term.
\begin{equation}\label{g2andg1}
G^{(2)}=G_{ent}^{(2)}+G_{A}^{(1)}G_{B}^{(1)}
\end{equation}
where $G_{A}^{(1)}$ and $G_{B}^{(1)}$ are the first order
correlation functions for the two arms. the The coincident
counting rate is then calculated using,
\begin{eqnarray}\label{rate}
R & \propto & \frac{1}{T}\int dt_{1}\int dt_{2}G^{(2)}(\vec{r}_{1},t_{1},\vec{r}_{2},t_{2})\nonumber\\
& = & \frac{1}{T}\int dt_{1}\int
dt_{2}[G_{ent}^{(2)}+G_{A}^{(1)}G_{B}^{(1)}]
\end{eqnarray}
To make the discussion easier, we shall evaluate the two terms in
Eq~(\ref{rate}) separately. First we make a few approximations. It
is usually easier to work with quantities invariant along the
beam: $\omega$, the angular frequency, and $\vec{q}$, the
component of the wave vector parallel to the output face of the
crystal. The $z$ component of the wave vector for a photon of
polarization $\beta$ is \vspace{-0.09in}
\begin{equation}
\label{kz} k_{z} = \sqrt{\left(\frac{\omega
n_{\beta}(\omega)}{c}\right)^{2}-q^{2}}
\end{equation}
Outside the crystal, the index of refraction $n_{\beta}=1$. Inside
the crystal, $n_{\beta}$ depends on the orientation of the optic
axis with respect to the wave vector of the beam~\cite{born}. We
assume $|\vec{q}|\ll|\vec{k}|$ and $k_{z}\cong \omega
n_{\beta}(\omega)/c$, so that \vspace{-0.09in}
\begin{equation}
\label{kz-approx} dk_{z} = \frac{d\omega}{u_{\beta}}
\end{equation}
where $u_{\beta}$ is the group velocity of a photon of
polarization $\beta$. Since the integrals in
Eq.~(\ref{f-corr-int}) are over modes outside the crystal,
\vspace{-0.09in}
\begin{equation}\label{dk-out}
\int d^{3}k \rightarrow  \int d^{2}q d\omega
\frac{\omega}{c^{2}k_{z}} \cong \int d^{2}q \frac{d\omega}{c}
\end{equation}

We make the following approximations to further simplify the
problem. We assume that the central frequencies of the signal and
idler are degenerate, ie \vspace{-0.09in}
\begin{eqnarray}\label{freq.}
    \omega_{o} & = & \Omega_{o} + \nu\\
    \omega_{e} & = & \Omega_{e} - \nu \\
    \Omega_{e} & \cong & \Omega_{o} \cong \omega_{p}/2
\end{eqnarray}
Further we will assume that $\nu \ll \Omega_{o,e}$ and $\omega_{o}
\cong \omega_{e} \cong \omega_{p}/2$ everywhere except the
exponential terms. The squeeze parameter $|\xi_{k_{s}}|$ is
considered constant for all modes. With these approximations, we
can now calculate the first term in the rate of coincidence
counting.
\begin{eqnarray}\label{first-term}
R_{ent} & = & \frac{1}{T}\int dt_{1}\int dt_{2}G_{ent}^{(2)}\\
& = & \frac{1}{T}\int _{0}^{T}dt_{1}\int _{0}^{T}dt_{2}
\left(\frac{1}{2\pi}\right)^{6} \left(\frac{\hbar
\omega_{p}}{4\epsilon_{0}c}\right)^{2} \nonumber
\\  & & \left|\int d^{2}q_{e}g_{A}(\frac{\omega_{p}}{2},
-\vec{q}_{e},0,d_{1}+d^{'}_{1}) g_{B}(\frac{\omega_{p}}{2},
\vec{q}_{e},\vec{\rho}_{2},d_{2})\right.\nonumber
\\  & &
\label{true-detail}
 \left.\int d\nu e^{-i\nu
(t_{1}-t_{2})}\right|^{2} cosh^{2}(|\xi|) sinh^{2}(|\xi|)
\end{eqnarray}
$R_{ent}$ represents 'true' coincidences of mutually entangled
photons. The Green's functions in Eq.~(\ref{true-detail})imply
that that the detector $D_{1}$ is fixed at the origin so that
$\vec{r}_{1} = (d_{1}+d^{'}_{1}) \hat{z}$ and the position of
detector $D_{2}$ is $\vec{r}_{2} =
\rho_{2}\hat{\rho}+d_{2}\hat{z}$(see fig 1). $T$ is the time
window of coincidence detection. If the two detectors register two
photons within time $T$ of each other, the photons are assumed to
be part of one entangled pair. For the present calculation $T =
1.8$ ns. We have also used the fact that for a given pair of
signal and idler modes related by the phase matching conditions,
$\vec{q}_{o} = -\vec{q}_{e}$. Since the visibility, being a ratio,
is not affected by the finite detection area of the detectors, we
will not bother about them in this calculation. We now look at the
frequency and time integrals in Eq.~(\ref{first-term}). If the
bandwidth of light reaching the two detectors, $\Delta$, is such
that $ \Delta T \gg 1$, we can approximate the frequency and time
integrals by $2 \pi \Delta T $~\cite{klyshko, rubin}.
\begin{eqnarray}\label{first-nogreen}
R_{ent} & = & 2 \pi \Delta \left(\frac{1}{2\pi}\right)^{6}
\left(\frac{\hbar
\omega_{p}}{4\epsilon_{0}c}\right)^{2}cosh^{2}(|\xi|)
sinh^{2}(|\xi|)\\
& & \left|\int d^{2}q_{e}g_{A}(\frac{\omega_{p}}{2},
-\vec{q}_{e},0,d_{1}+d^{'}_{1}) g_{B}(\frac{\omega_{p}}{2},
\vec{q}_{e},\vec{\rho}_{2},d_{2})\right|^{2}\nonumber
\end{eqnarray}

 Using the
Green's functions from the appendix A, $R_{ent}$ (ignoring common
constants) is given by,
\begin{eqnarray}\label{first-term-2}
R_{ent}& = & (2\pi\Delta)\left(\frac{\omega_{p}}{2c}\right)^{4}
\left(\frac{1}{d^{'}_{1}(d_{1}+d_{2})}\right)^{2}cosh^{2}(|\xi|)
\nonumber\\
& & sinh^{2}(|\xi|)\times
|\tilde{t}(\frac{\omega_{p}}{2c}\frac{\vec{\rho}_{2}}{d_{1}+d_{2}})|^{2}
\end{eqnarray}
where $\tilde{t}(\vec{q})$ is the fourier transform of the
aperture function in wave vector space. So for a double-slit
aperture we have the expected 'Ghost'interference pattern in the
true coincidences. $cosh^{2}(|\xi|)sinh^{2}(|\xi|)$ is the
amplification factor. Now we look at the accidental coincidence
term,
\begin{eqnarray}\label{acc-coincidence}
R_{acc} & = & \frac{1}{T}\int dt_{1}\int dt_{2}G_{A}^{(1)}G_{B}^{(1)}\\
 & = & \frac{1}{T}\int _{0}^{T}dt_{1}\int _{0}^{T}dt_{2} \left(\frac{1}{2\pi}\right)^{6}
\left(\frac{\hbar
\omega_{p}}{4\epsilon_{0}c}\right)^{2}sinh^{4}(|\xi|)
\nonumber \\
& & \left[\int d\omega^{'}_{o}\int
d^{2}q^{'}_{o}|g_{A}(\frac{\omega_{p}}{2},
\vec{q}^{'}_{o},0,d_{1}+d^{'}_{1})|^{2}\right.\nonumber
\\& & \left.\int d\omega_{e}\int
d^{2}q_{e}|g_{B}(\frac{\omega_{p}}{2}, \label{acc-detail}
\vec{q}_{e},\vec{\rho}_{2},d_{2})|^{2}\right]
\end{eqnarray}
Completing the time and frequency integrals in
Eq.~(\ref{acc-detail}), we can now infer the effect of the
coincidence window $T$ in a general case without entering into the
details of the experiment.
\begin{eqnarray} \label{acc-window}
R_{acc}& = & T \left(\frac{1}{2\pi}\right)^{6}\left(\frac{\hbar
\omega_{p}}{4\epsilon_{0}c}\right)^{2}\times \nonumber \\
& & \left[\Delta sinh^{2}(|\xi|)\int
d^{2}q^{'}_{o}|g_{A}(\frac{\omega_{p}}{2},\vec{q}^{'}_{o},0,d_{1}+d^{'}_{1})|^{2}\right.\nonumber
\\ & &  \left.\times\Delta sinh^{2}(|\xi|)\int
d^{2}q_{e}|g_{B}(\frac{\omega_{p}}{2},\vec{q}_{e},\vec{\rho}_{2},d_{2})|^{2}\right]
\end{eqnarray}
Comparing $R_{ent}$ and $R_{acc}$ from Eq's.~(\ref{first-nogreen})
and (\ref{acc-window}), we find that at low gain, $R_{acc}$ is
negligible compared to $R_{ent}$ since
$sinh^{4}(|\xi|)\rightarrow(|\xi|)^{4}$. The coincidence window
has no effect on the result as only a single pair of photons is
produced within the time $T$. But as the gain increases, and more
pairs of photons are produced within  a given time interval,
$R_{acc}$ becomes significant and the visibility of the
interference pattern begins to fall.

Using the Green's functions and a double-slit with aperture
function,
\begin{equation}\label{aperture}
    t(\vec{\rho}_{a}) = \left[rect\left(\frac{x+a/2}{b}\right)+
    rect\left(\frac{x-a/2}{b}\right)\right]rect\left(\frac{y}{B}\right)
\end{equation}
where $a$ and $b$ are the distance between the centers of the slit
and width of the slits respectively and $B$ is the length of the
slit, $R_{acc}$(ignoring common constants) for the
'Ghost'interference experiment is found to be
\begin{eqnarray}\label{acc-final}
R_{acc}& = & \Delta^{2} T \left(\frac{\omega_{p}}{2 \pi c
d^{'}_{1}}\right)^{2}sinh^{4}(|\xi|)\int d^{2}
q^{'}_{o}|\tilde{t}(\vec{q}^{'}_{o})|^{2} \nonumber \\ & & \times
\int
d^{2}q_{e}|g_{B}(\omega_{p}/2,\vec{q}_{e},\vec{\rho}_{2},d_{2})|^{2}
\end{eqnarray}
For a given gain value $|\xi|$ and coincidence window $T$,
$R_{acc}$ is a constant since the interference pattern behind the
double-slit in arm A (see Fig.~1) is averaged over due to the
bandwidth of the source and the intensity at the scanning
detector, uniformly illuminated by the SPDC beam, is a constant.
The details of this calculation are given in appendix B.

The complete expression of the coincidence counting rate in the
'Ghost' interference pattern is given by
\begin{eqnarray}\label{total-final}
R & \propto & R_{ent}+R_{acc} \\
& = & (2\pi\Delta)\left(\frac{ \omega_{p}}{2c}\right)^{4}
\left(\frac{Bb}{d^{'}_{1}(d_{1}+d_{2})}\right)^{2}cosh^{2}(|\xi|)\nonumber\\
 & & sinh^{2}(|\xi|)\left\{
Sinc^{2}\left(\frac{\omega_{p}}{2c}\frac{\rho_{2x}b}{2(d_{1}+d_{2})}\right)
\times \right. \nonumber\\  & &
\left.Cos^{2}\left(\frac{\omega_{p}}{2c}\frac{\rho_{2x}a}{2(d_{1}+d_{2})}\right)
Sinc^{2}\left(\frac{\omega_{p}}{2c}\frac{\rho_{2y}B}{2(d_{1}+d_{2})}\right)\right\}
  \nonumber\\  & &  +\Delta^{2} T \left(\frac{\omega_{p}}{2 \pi c
d^{'}_{1}}\right)^{2}sinh^{4}(|\xi|)\int d^{2}
q^{'}_{o}|\tilde{t}(\vec{q}^{'}_{o})|^{2}\nonumber\\  & & \int
d^{2}q_{e}|g_{B}(\omega_{p}/2,\vec{q}_{e},\vec{\rho}_{2},d_{2})|^{2}
\end{eqnarray}
\begin{figure}[h]
\label{visibility}
\includegraphics [width=3.4in]{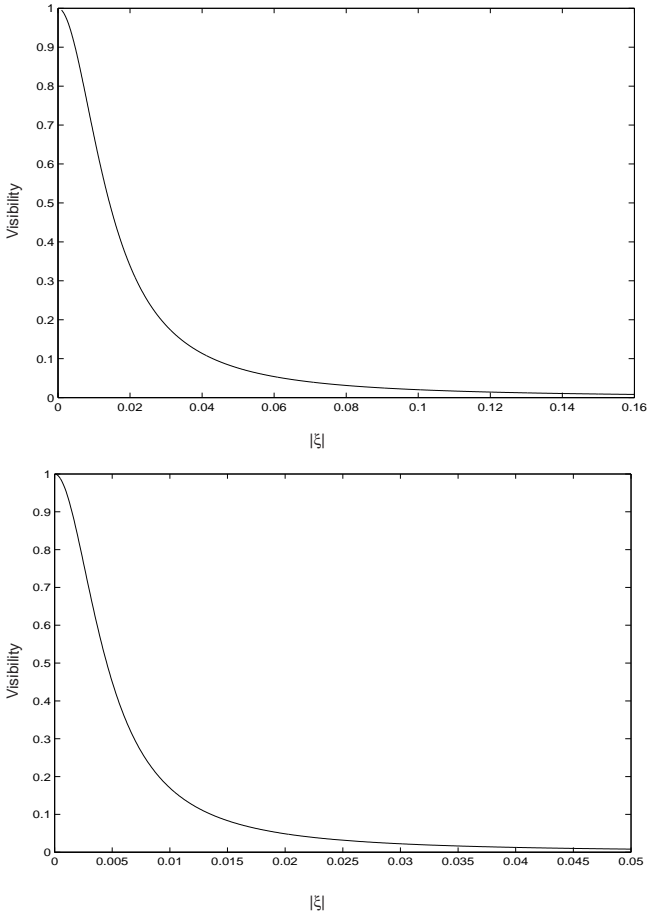}
\caption{\protect The visibility of interference fringes as a
function of parametric gain, when the coincidence time window $T =
1.8$ns and bandwidth of the light reaching the detectors is (a)1nm
and (b)10nm}
\end{figure}
The visibility of the interference pattern as a function of the
parametric gain, calculated from the expression for coincidence
count rate in Eq.~(\ref{total-final}), is  given by
\begin{equation}\label{visibility}
    V = \frac{A~cosh^{2}(|\xi|)}{A~cosh^{2}(|\xi|)+ 2 ~B ~(\Delta T) ~sinh^{2}(|\xi|)}
\end{equation}
where $A$ and $B$ are constants arising from experimental factors.
Fig.~2 shows plots of visibility as the bandwidth of light
reaching the detectors is increased.

We now look into the features of the visibility plots and analyze
the factors that give rise to these features. The important
parameters for this purpose are the gain terms $cosh(|\xi|)$,
$sinh(|\xi|)$ and $\Delta T$ in Eq.~(\ref{total-final}). The
bandwidth $\Delta$ is a measure of the number of modes since we
have assumed perfect phase matching. $\Delta T$ can be thought of
as a measure of the detectors' ability to resolve two entangled
pairs of photons. A  value of $\Delta T\gg1$ leads to increase in
accidental counts since detectors cannot distinguish every
entangled pair of photons. The effect of the coincidence window
time $T$ is easy to understand. In the very low gain limit
($sinh|\xi|)\approx 0$) only one single pair of entangled photons
is produced within time $T$ and so every entangled pair can be
distinguished. As the gain increases, many more entangled pairs
are produced and reach the detectors within the coincidence window
time causing accidental coincidences of photons belonging to
different entangled pairs, and hence visibility is lost.

To understand the effect of the number of modes on the visibility,
we go back to the output state of the OPA. The OPA state is
restricted to include the two photon state and the next higher
order interaction giving four photon states.

The output at the OPA, given in Eq.~(\ref{output-state}) can be
written as
\begin{equation}\label{opa-out}
|\psi\rangle = \prod_{\vec{k}_{s}} exp[\frac{-\xi_{k_{s}}}{2}
a^{\dagger}_{\vec{k}_{s}}(0)a^{\dagger}_{\vec{k}_{i}}(0)]|0\rangle
\end{equation}
where $\vec{k}_{s}+\vec{k}_{i}=\vec{k}_{p}$ is the phase matching
condition between the signal, idler and pump modes. We consider
$n$ pairs of signal and idler modes, expand the exponential
operator term in Eq.~(\ref{opa-out}) and omit  terms greater than
second order in the gain parameter $\xi$. If the gain is
considered constant for all the modes, then the unnormalized OPA
state is given by
\begin{eqnarray}\label{state-normal}
|\psi\rangle & = & |0\rangle
-\left(\frac{|\xi|}{2}\right)\sum_{\vec{k}_{s}} |1_{\vec{k}_{s}}
1_{\vec{k}_{i}}\rangle +  \\\nonumber & &
\left(\frac{|\xi|^{2}}{8}\right)\left[
\sum_{\vec{k}_{s}}|2_{\vec{k}_{s}} 2_{\vec{k}_{s}}\rangle +
\sum_{\vec{k}_{s},\vec{k}^{'}_{s}} |1_{\vec{k}_{s}}
1_{\vec{k}_{i}}\rangle|1_{\vec{k}^{'}_{s}}
1_{\vec{k}^{'}_{i}}\rangle\right]
\end{eqnarray}
From the expression for the truncated OPA state in
Eq.~(\ref{state-normal}) we infer that the states in the first and
second terms of the equation lead to 'good' coincidence counts.
The detectors detect photons belonging to a single entangled pair
or to an entangled four photon state. The third term, on the other
hand, can lead to detection of two photons belonging to different
entangled states ('bad' counts) half of the time. The conditional
probability of getting a good count, given $n$ modes, is found to
be
\begin{equation}\label{prob}
    P(good|n) = \frac{16+(\frac{n+1}{2}) \xi^{2}}{16+n \xi^{2}}
\end{equation}

For a large number of modes, this probability tends to 0.5. This
implies that when a large number of modes are allowed, the
visibility of the coincidence detection pattern falls to 0.5.

Though the number of good counts seem to dominate according to
Eq.~(\ref{prob}), the number of modes along with the coincidence
window time produce a loss in visibility greater than 0.5. As the
gain increases the probability of both good and bad counts
increase and tend towards a constant limit. This leads to the
flattening of the visibility with rising parametric gain. The
flattening of the visibility occurs at lower gain as the
bandwidth(and number of modes)increases. The value of the limiting
visibility falls as the number of allowed modes increases. This is
to be expected since as the gain $\xi$ rises, entangled pairs (the
first order states of the OPA) are emitted in all modes. If the
number of modes allowed in the experiment are increased then the
number of possible good and bad counts also increase. The
coincidence time window $T$ further adds to the bad counts causing
a greater fall in visibility and the limiting visibility is
lowered.

If a smaller number of modes are allowed into the experiment, for
example by using a fine pinhole, the visibility can be maintained
at values greater than 0.5 for larger values of gain. But even in
the case of just two modes, the visibility eventually falls due to
the higher order states of the OPA and tends to a constant as
$cosh(|\xi|)\approx sinh(|\xi|) \approx e^{|\xi|}/2$.

This is the fundamental difference in the mechanism of visibility
loss with two-mode and multimode OPA. In a multimode OPA the loss
of visibility is mainly due to the number of modes, and occurs at
much lower gain than the two-mode OPA, where the higher order
terms lead to loss of visibility for a given coincidence window
time.

\section{Discussion and Summary}
We have analyzed the effect of a multimode optical parametric
amplifier source in the 'Ghost' interference experiment, taking
into account the finite coincidence window of the photon counting
detectors. We find that the loss of visibility with increasing
parametric gain is strongly dependent on the coincidence time
window. A longer coincidence time window reduces the ability of
the photon counting detectors to resolve entangled pairs as the
parametric gain of the OPA increases. We have also highlighted the
differences between effects observed with a two-mode and a
multimode source. An increase in the number of modes in the
experiment increases the probability of accidental coincidences
between photons belonging to different entangled pairs (or
states). Further, this experiment limited loss of visibility
occurs in all coincidence counting measurements though it is
significant only in the regime of a strong source of entangled
photons like an OPA. We conclude that a cautious choice of sources
and detection schemes are necessary in order to observe certain
signatures of entangled light in a macroscopic regime.

Since the completion of this calculation we have become aware of a
two-photon absorption technique demonstrated by Dayan et.
al.~\cite{Dayan} where Rb atoms undergoing simultaneous absorption
of signal and idler photons overcome the problem of temporal
resolution associated with a  strong broadband source like the
multimode OPA. The two-photon transition is sensitive to minute
delays (order of 100 fs) between the signal and idler photons. But
such a detection scheme does not discriminate between entangled
and separable pairs of photons and cannot reduce the loss of
visibility in a 'Ghost' interference experiment. Further, the use
of such highly sensitive detection scheme requires a high gain OPA
source which enhances accidental coincidence counts.

\acknowledgements
 The authors would like to thank their colleagues from the UMBC
 Quantum Optics Group, M. D'Angelo, A. Valencia,
 G. Scarcelli, J. Wen and Y. H. Shih, for discussions about the
 material in this paper. This work was supported in part by NSF
 grant OSPA 2001-0176
\appendix
\section{Green's Functions for propagation through a linear optical system}
We give a brief review of propagation of an electric field through
a diffraction limited linear optical system in each arm of the
experimental set-up, following the treatment in ~\cite{rubin}. The
positive frequency part of the electric field at a time $t$ at the
input of a detector at $\vec{r} = z \hat{z}+ \vec{\rho}$ is given
by
\begin{equation}\label{E-feild}
    E_{\beta}^{(+)} = \int\int d\omega d^{2}\vec{q}E(\omega)
    g(\vec{q},\omega,\vec{\rho},z) a_{\beta}(\omega, \vec{q})
\end{equation}
where $ a_{\beta}(\omega, \vec{q})$ is the annihilation operator
at the source for a photon of angular frequency $\omega$,
transverse wave vector $\vec{q}$ and polarization $\beta$. The
unit vector $\hat{z}$ is the inward normal to the detector
surface. $E(\omega)$ is a slowly varying function required for
dimensional reasons and can be assumed constant in the current
analysis. $ g(\vec{q},\omega,\vec{\rho},z)$ is the optical
transfer function or Green's function which describes propagation
through the linear optical system. In classical electromagnetic
theory $g$ connects electric fields in real space. In our quantum
mechanical analysis, it connects two operators in photon number or
Fock space. The superposition principles involved in calculating
$g$ are purely classical from classical electromagnetic theory. So
in the quantum mechanical context it is best thought of as arising
from boundary conditions on the modes of the fields irrespective
of the state of the system.

We now calculate the Green's function for the arm A in the
Fig.~(1). The Green's function is expressed in terms of the
aperture function defined by $t(\vec{\rho}_{a})$. \vspace{-0.15in}
\begin{equation}\label{g-a}
\begin{array}{l}
    g_{A}(\vec{q},\omega,0,d_{1}+d^{'}_{1}) = \\
    \int d^{2}\rho_{a}
    \int d^{2} \rho_{s} h_{\omega}(-\vec{\rho}_{a}, d^{'}_{1})
    t(\vec{\rho}_{a}) h_{\omega}(\vec{\rho}_{a}-\vec{\rho}_{s},
    d_{1})e^{i \vec{q}\cdot \vec{\rho}_{s}}
\end{array}
\end{equation}
where $\vec{\rho}_{s}$ and $\vec{\rho}_{a}$ are the transverse
co-ordinates of the source (crystal) plane and aperture plane. In
the Fresnel approximation~\cite{goodman},
\begin{eqnarray}\label{fresnel}
    h_{\omega}(\vec{\rho}, d)& = & (\frac{-i\omega}{2\pi
    c})\frac{e^{i(\omega/c)d}}{d}
    \psi(|\vec{\rho}|,\frac{\omega}{cd}) \\
\psi(|\vec{\rho}|,\frac{\omega}{cd})& = &
e^{i(\omega/2cd)\rho^{2}}
\end{eqnarray}
Finally \vspace{-0.15in}
\begin{equation}\label{g-a-final}
\begin{array}{l}
    g_{A}(\vec{q},\omega,0,d_{1}+d^{'}_{1}) = (\frac{-i\omega}{2\pi c})
    \frac{e^{i(\omega/c)(d_{1}+d^{'}_{1})}}{d^{'}_{1}}  \\
    \times \int
    d^{2}\rho_{a} \psi(|\vec{\rho}_{a}|,\frac{\omega}{cd^{'}_{1}})t(\vec{\rho}_{a})
    e^{i \vec{q}\cdot \vec{\rho}_{a}}  \psi(|\vec{q}|,-\frac{c}{\omega}d_{1})
\end{array}
\end{equation}

The Green's function in the arm B of the experimental set-up, for
each plane wave mode, assuming that the source has a large
cross-section is given by
\begin{eqnarray}\label{g-b}
g_{B}(\vec{q},\omega,\vec{\rho},d_{2}) = \int d^{2}
\rho_{s}h_{\omega}(\vec{\rho}-\vec{\rho}_{s},
    d_{2})e^{i \vec{q}\cdot \vec{\rho}_{s}} \\
    = e^{i(\omega/c)d_{2}} e^{i \vec{q}\cdot \vec{\rho}}
    \psi(|\vec{q}|,-\frac{c}{\omega}d_{2})
\end{eqnarray}
Using the above expressions for $g_{A}$ and $g_{B}$, we see that,
\begin{equation}\label{gagb}
\begin{array}{l}
\int d^{2}q
g_{A}(-\vec{q},\omega,0,d_{1}+d^{'}_{1})g_{B}(\vec{q},\omega,\vec{\rho},d_{2})
= \\
(\frac{-i\omega}{2\pi c})
    \frac{e^{i(\omega/c)(d_{1}+d^{'}_{1}+d_{2})}}{d^{'}_{1}}\int
    d^{2}\rho_{a} \psi(|\vec{\rho}_{a}|,\frac{\omega}{cd^{'}_{1}})t(\vec{\rho}_{a})\\
    \times \int d^{2}q e^{i \vec{q}\cdot
    (\vec{\rho}-\vec{\rho}_{a})}\psi(|\vec{q}|,-\frac{c}{\omega}(d_{1}+d_{2})) \\
    =
-\left(\frac{\omega}{c}\right)^{2}\frac{e^{i(\omega/c)(d_{1}+d^{'}_{1}+d_{2})}}
    {d^{'}_{1}(d_{1}+d_{2})}
    \psi(|\vec{\rho}|,\frac{\omega}{c}\frac{1}{d_{1}+d_{2}}) \\
    \times \int
    d^{2}\rho_{a} \psi(|\vec{\rho}_{a}|,\frac{\omega}{c}(\frac{1}{d^{'}_{1}}+\frac{1}{d_{1}+d_{2}}))
    e^{-i\frac{\omega}{c}\frac{1}{d_{1}+d_{2}}\vec{\rho}\cdot\vec{\rho}_{a}} t(\vec{\rho}_{a})
    \end{array}
\end{equation}
In the far field Fraunhofer approximation, the $\psi$'s in the
above expression go to unity and
\begin{equation}\label{intqgagb}
\begin{array}{l}
\int d^{2}q
g_{A}(-\vec{q},\omega,0,d_{1}+d^{'}_{1})g_{B}(\vec{q},\omega,\vec{\rho},d_{2})=
\\
 -\left(\frac{\omega}{c}\right)^{2}\frac{e^{i(\omega/c)
(d_{1}+d^{'}_{1}+d_{2})}}{d^{'}_{1}(d_{1}+d_{2})}
(2\pi)\tilde{t}(\frac{\omega}{c}\frac{\vec{\rho}}{d_{1}+d_{2}})
\end{array}
\end{equation}
where $\tilde{t}$ is the fourier transform of the aperture
function. Further, \vspace{-0.16in}
\begin{equation}\label{mod-squared}
\begin{array}{l}
   |g_{A}|^{2} = \\
   -(\frac{i\omega}{2\pi c})^{2}
   \frac{1}{d^{'2}_{1}} \left[\int
    d^{2}\rho_{a} \psi(|\vec{\rho}_{a}|,\frac{\omega}{cd^{'}_{1}})t(\vec{\rho}_{a})
    e^{i \vec{q}\cdot \vec{\rho}_{a}}
    \psi(|\vec{q}|,-\frac{c}{\omega}d_{1})\right] \\
    \times \left[\int
    d^{2}\rho^{'}_{a} \psi(|\vec{\rho}^{'}_{a}|,\frac{\omega}{cd^{'}_{1}})t(\vec{\rho}^{'}_{a})
    e^{i \vec{q}\cdot \vec{\rho}^{'}_{a}}
    \psi(|\vec{q}|,-\frac{c}{\omega}d_{1})\right]^{\ast}
\end{array}
\end{equation}
In the far- field Fraunhofer approximation,
\begin{eqnarray}\label{mod-squared-fraunhofer}
   |g_{A}|^{2} & = & (\frac{\omega}{2\pi c})^{2}
   \frac{1}{d^{'2}_{1}} |\int d^{2}\rho_{a}t(\vec{\rho}_{a})e^{i \vec{q}\cdot
   \vec{\rho}_{a}}|^{2} \\
   & = & (\frac{\omega}{2\pi
   c})^{2}\frac{1}{d^{'2}_{1}}(2\pi)^{2}|\tilde{t}(\vec{q})|^{2}
\end{eqnarray}

\section{Singles Detection} We now look at results of detection
at the detectors $D_{1}$ and $D_{2}$ individually, without caring
for coincidences. This involves calculating the first order
correlation functions~\cite{glauber},
\begin{equation}\label{G1}
    G^{(1)}(\vec{r}_{i}, t_{i}) = \langle0|E_{i}^{(-)}(\vec{r}_{i}, t_{i})
    E_{i}^{(+)}(\vec{r}_{i}, t_{i})|0\rangle
\end{equation}
where $i=1,2$. Using the same techniques as in coincidence
detection, we find that the first order correlation functions are
given by
\begin{eqnarray}\label{G1s}
G_{A}^{(1)}(\vec{r}_{1}, t_{1}) & = &
\left(\frac{1}{2\pi}\right)^{3} \left(\frac{\omega_{p}}{4\pi c
d^{'2}_{1}}\right)^{2}
\left(\frac{\hbar\omega_{p}}{4\epsilon_{0}c}\right) \\ \nonumber &
& \times sinh^{2}(|\xi|)\Delta \int
d^{2}q_{o}|\tilde{t}(q_{o})|^{2} \\
G_{B}^{(1)}(\vec{r}_{2}, t_{2}) & = &
\left(\frac{1}{2\pi}\right)^{3}
\left(\frac{\hbar\omega_{p}}{4\epsilon_{0}c}\right)\Delta
sinh^{2}(|\xi|)\\ \nonumber & & \times \int
d^{2}q_{e}|g_{B}(\omega_{p}/2,\vec{q}_{e},\vec{\rho}_{2},d_{2})|^{2}
\end{eqnarray}
$G_{B}^{(1)}(\vec{r}_{2}, t_{2})$ is the correlation function in
the arm without the double slit and as expected, there is no
interference pattern at $D_{2}$. The correlation function in the
arm with the double-slit $G_{A}^{(1)}(\vec{r}_{1}, t_{1})$,
suggests that there might be a pattern at $D_{1}$, if the
transverse components of the wave vector ($q_{o}$) reaching the
detector is narrow enough. For the wavelengths and the dimensions
of double-slit chosen, an interference pattern will be observed if
the detection angle of the SPDC beam at the detector $\Delta
\theta_{\perp}< 2\lambda_{p}/a \approx 1.5$ mrad. But for the 1nm
filter used, $\Delta \theta_{\perp}\approx 15$ mrad. Therefore, no
interference pattern is found behind the double slit due to the
large divergence of the SPDC beam.

\end{document}